# From Protoplanetary Disks to Extrasolar Planets: Understanding the Life Cycle of Circumstellar Gas with Ultraviolet Spectroscopy

Kevin France[1*], Matthew Beasley[1], David R. Ardila[2], Edwin A. Bergin[3], Alexander Brown[1], Eric B. Burgh[1], Nuria Calvet[3], Eugene Chiang[4], Timothy A. Cook[5], Jean-Michel Désert[6], Dennis Ebbets[7], Cynthia S. Froning[1], James C. Green[1], Lynne A. Hillenbrand[2], Christopher M. Johns-Krull[8], Tommi T. Koskinen[9], Jeffrey L. Linsky[1], Seth Redfield[10], Aki Roberge[11], Eric R. Schindhelm[12], Paul A. Scowen[13], Karl R. Stapelfeldt[11], and Jason Tumlinson[14]

[1]University of Colorado, [2]Caltech, [3]University of Michigan, [4]University of California, Berkeley, [5]Univserity of Massachusetts, Lowell, [6]Harvard/CfA, [7]Ball Aerospace, [8]Rice University, [9]University of Arizona, [10]Wesleyan University, [11]NASA/GSFC, [12]SwRI, [13]Arizona State University, [14]STScI

1. **Introduction**

Few scientific discoveries have captured the public imagination like the explosion of exoplanetary science during the past two decades. This work has fundamentally changed our picture of Earth's place in the Universe and led NASA to make significant investments towards understanding the demographics of exoplanetary systems and the conditions that lead to their formation. The story of the formation and evolution of exoplanetary systems is essentially the story of the circumstellar gas and dust that are initially present in the protostellar environment; in order to understand the variety of planetary systems observed, we need to understand the life cycle of circumstellar gas from its initial conditions in protoplanetary disks to its endpoint as planets and their atmospheres. In this white paper response to NASA's Request for Information "*Science Objectives and Requirements for the Next NASA UV/Visible Astrophysics Mission Concepts (NNH12ZDA008L)*", we describe scientific programs that would use the unique capabilities of a future NASA ultraviolet (UV)/visible space observatory to make order-of-magnitude advances in our understanding of the life cycle of circumstellar gas.

UV radiation plays a critical role in the evolution of protoplanetary disks, the heating and evaporation of extrasolar planets, and directly probes the most abundant molecules in these environments. We outline four broad scientific investigations that address these topics using UV spectral observations. We first describe *1)* the importance of UV observations in understanding the production of biomarkers on potentially habitable, Earth-like planets, and *2)* the characterization of exoplanetary atmospheres, using spectroscopy to probe the compositions and thermodynamic structures of transiting planets. The Astro2010 Decadal Survey lists **"How do circumstellar disks evolve and form planetary systems?"** as a Frontier Science question for Cosmic Origins in the present decade (Blandford et al. 2010). We propose that this can be addressed with *3)* high-resolution UV molecular spectroscopy to measure the structure and composition of the disk and *4)* wide-field UV surveys of molecular emission that will allow statistical determination of protoplanetary gas disk lifetimes and the implications for the formation and evolution of exoplanetary systems.

[*]kevin.france@colorado.edu; 1-303-492-1429

## 2. Characterization of Exoplanet Atmospheres: Terrestrial Worlds and Gas Giants

### 2.1 The Habitable Zone around Low-Mass Stars

The ultimate goal for exoplanetary science in the next two decades is the detection and characterization of habitable, Earth-like worlds. Stellar characterization is critical to interpreting these observations. An investment in stellar characterization is particularly important for low-mass stars (M- and K-dwarfs), which are perhaps the most promising targets for the detection of habitable planets (Segura et al. 2005; Rauer et al. 2011). For example, UV radiation is important to the photodissociation and photochemistry of $H_2O$ and $CO_2$ in terrestrial planet atmospheres. The large far-UV/near-UV stellar flux ratio in the habitable zone around M-dwarfs can have a profound influence on the atmospheric oxygen chemistry on Earth-like planets. The strong far-UV flux may produce large, abiotic atmospheric abundances of $O_2$ through the dissociation of $CO_2$ (France et al. 2012). The $O_2$ production rate and the subsequent formation of $O_3$ are highly dependent on the spectral and temporal behavior of the far-UV (in particular, Ly$\alpha$) and near-UV radiation field of the host star. Over 100 extrasolar planets (or planet candidates) orbiting M-dwarfs are known today, however only *three* of these systems have measured far- and near-UV spectra. At present, we cannot accurately predict the UV spectrum of an M-star; this lack of observational basis is hampering our ability to accurately predict the expected biosignatures from these worlds (Kaltenegger et al. 2011).

A spectroscopic survey of low-mass exoplanet host stars from 912 – 4000 Å will be able to characterize the spectral and temporal behavior of these systems, an essential input for atmospheric models of habitable zone planets. With *HST*, we can only observe the UV spectrum of an M-dwarf exoplanet host out to $d \sim 10$ pc with a reasonable investment in observing time (< 8 orbits), therefore if we want to determine the potential habitability of more than the nearest few Earth-like planets, a new observational capability is required. A future UV mission employing **moderate resolution (~10 km s$^{-1}$), low-background equivalent flux levels ($\leq 10^{-18}$ erg cm$^{-2}$ s$^{-1}$ Å$^{-1}$ in $10^4$ sec), and photon-counting detectors ($\Delta t \leq 1$ sec)** would enable a survey of the known M-dwarf exoplanetary host stars within 50 pc (and K-dwarfs to > 200 pc), including all of the systems that can be studied in detail by *JWST*. The ultraviolet bandpass offers the best set of chromospheric, transition region, and coronal activity diagnostics in low-mass stars (the HI Lyman series, FeXVIII 974, CIII $\lambda$977, OVI $\lambda$1032, SiIII $\lambda$1206, OI $\lambda$1304, CII $\lambda$1335, FeXXI 1354, CIV $\lambda$1550, HeII 1640 Å, and MgII 2800 Å) that are critical to the characterization of the energetic radiation environment. This work is an essential investment towards our ability to reliably interpret biosignature molecules when they are discovered in the coming decades.

### 2.2 Transiting Planets

Short period planets are exposed to strong UV radiation fields from their host stars, and this energy deposition can inflate the planetary atmosphere. UV observations probe the extended upper atmospheres of the planets, providing unique access to the strong resonant transitions of the most abundant atomic constituents that can be observed in absorption during transit (e.g., H, O, $C^+$, $Si^{2+}$, and $Mg^+$ have been detected so far; Vidal-Madjar et al. 2003, 2004, Linsky et al. 2010, Fossati et al. 2010).

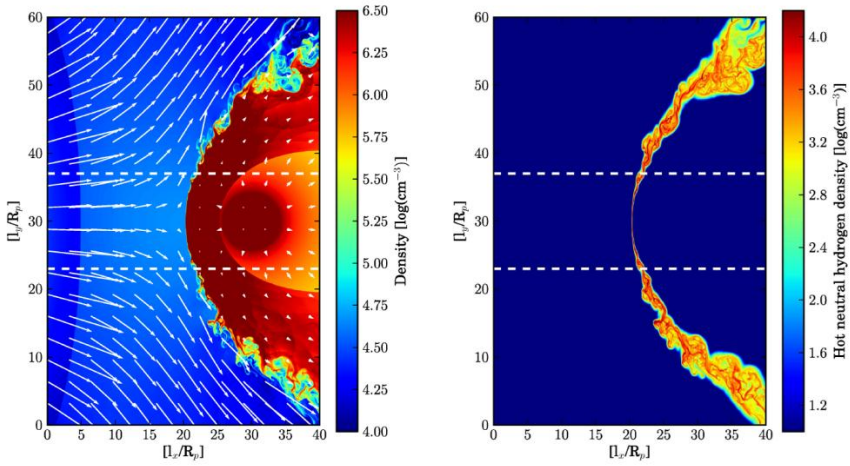

Figure 1: Hydrodynamic simulation of the interaction of stellar winds and the escaping neutral hydrogen atmosphere of a hot Jupiter (from Tremblin & Chiang 2012). UV transit observations provide a direct means to study the composition, structure, and evolution of exoplanetary atmospheres.

Together with models of the upper atmosphere, the observations can be used to study the ionization, chemistry, elemental abundances, thermal structures, and mass loss on transiting planets (e.g., Yelle 2004, Garcia Munoz 2007, Murray-Clay et al. 2009). Existing observations of close-in gas giant planets indicate that they are surrounded by thick envelopes of hot atomic hydrogen and ions that are created by photoionization and interaction with the stellar wind (see Figure 1 for an example). However, there are many competing interpretations of the observations (e.g., Vidal-Madjar et al. 2003, 2004, Ben-Jaffel and Hosseini 2010, Koskinen et al. 2010, Ekenback et al. 2010), and the combination of sensitivity and spectral resolution of the instruments aboard the *Hubble Space Telescope* are insufficient to resolve the differences between these interpretations. **Higher sensitivity ($\leq 10^{-17}$ erg cm$^{-2}$ s$^{-1}$ Å$^{-1}$ in $10^4$ sec) and resolution ($\Delta v < 3$ km s$^{-1}$) are required to study the dynamics of mass loss in extrasolar systems;** one of the primary motivations for future UV studies. Our current understanding of evaporating exoplanetary atmospheres are based on observations of *three planets*. Analogous to what we have learned from *Kepler*, when a statistical sample replaces a small number of easily observable targets, our fundamental understanding of exoplanetary atmospheres will almost certainly change. A future high sensitivity UV observatory is therefore necessary to carry out statistical surveys of transiting systems, including observations of planets orbiting lower mass stars, out to ~100 pc.

The primary constituent of gas giant atmospheres, molecular hydrogen ($H_2$), is best studied in the 912 – 1650 Å bandpass. $H_2$ lines are independent of the chromospheric variability that complicates UV transit studies of G, K, and M-stars; its narrow absorption lines will serve as excellent tracers of the temperature, molecular fraction, and velocity field of gas giant atmospheres. These lines could be detected in absorption against the bright CIII (977 Å), Lyβ (1026 Å), OVI (1032 Å), CIII (1175 Å), and Lyα (1216 Å) emission lines with high spectral resolution. The signatures of disintegrating rocky planets may also be observable against chromospheric metal lines (Rappaport et al. 2012).

In order to probe the details of atmospheric escape from "hot Jupiters", and eventually terrestrial mass planets, a new observational capability is required. Observations of Rayleigh scattering are the most direct means of determining the atmospheric scale height for both Jovian and terrestrial planets (Lecavelier des Etangs et al. 2008; Benneke & Seager 2012), an essential parameter for the interpretation of near- and mid-IR molecular transmission spectra from future (or proposed) NASA missions such as *FINESSE* and *JWST*. High-sensitivity, moderate spectral resolution **near-UV (1700 – 4000 Å) spectroscopy** would allow us to observe Rayleigh scattering of $H_2$, haze, and possibly $CO_2$ and $N_2$ atmospheres at the wavelengths where this mechanism has the largest observable signature (Sing et al. 2011).

3. The Structures, Compositions, and Lifetimes of Circumstellar Gas Disks

The lifetime, spatial distribution, and composition of gas and dust in the inner ~ 10 AU of young (age ≤ 30 Myr) circumstellar disks are important components for understanding the formation and evolution of extrasolar planetary systems. The formation of giant planet cores and their accretion of gaseous envelopes occurs on timescales similar to the lifetimes of the disks around T Tauri and Herbig Ae/Be stars ($10^6 – 10^7$ yr). The formation of giant planet cores through the coagulation of dust grains (Hayashi et al. 1985) is thought to be complete on the 2 – 4 Myr dust disk clearing timescale (Hernández et al. 2007). However, recent results indicate that inner molecular disks can persist to ages ~10 Myr in Classical T Tauri Stars (CTTSs, Salyk et al. 2009; Ingleby et al. 2011a; France et al. 2012b), although these results are based on a small number of protoplanetary systems. Disk gas regulates planetary migration (Ward 1997; Armitage et al. 2002; Trilling et al. 2002) and therefore the migration timescale is sensitive to the specifics of the disk surface density distribution and dissipation timescale (Armitage 2007). Below, we describe two experiments that would observationally constrain the structures, compositions, and lifetimes of circumstellar gas disks; allowing us to better understand the formation and evolution of exoplanetary systems.

### 3.1 Measuring the Radial Structures and Elemental Abundances of Gas Disks

At the distances of typical star-forming regions (e.g., Taurus-Auriga or the Orion Nebula Cluster), 1 AU corresponds to an angular scale of < 0.01´´. ALMA is carrying out high-resolution molecular spectroscopy of protoplanetary disks, but is less sensitive to warm/hot gas at terrestrial planet-forming radii. Therefore, if one wishes to probe molecules in the region of terrestrial and giant planet-formation, UV and IR spectroscopy will be the technique of choice for the foreseeable future. UV spectroscopy is

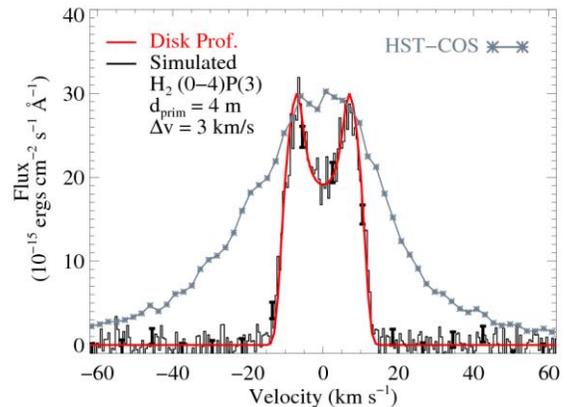

Figure 2: High-resolution ($\Delta v$ = 3 km s$^{-1}$) $H_2$ line profiles can be modeled to constrain radial distribution of the molecular disk. A high-throughput echelle spectrograph enables detailed (S/N ≈ 20 in 3000 sec) mapping of the molecular disk from $r$ ~ 0.1 – 10 AU for moderate inclination targets.

a unique tool for observing the inner molecular disk as **the strongest electronic band systems of $H_2$ and CO reside in the 1000 – 1700 Å bandpass** (Herczeg et al. 2002; France et al. 2011). We currently lack the combination of sensitivity and spectral resolution to measure the relative contributions of thermal, turbulent, and kinematic broadening of the molecular lines at planet-forming radii. **High-resolution ($\Delta v$ < 3 km s$^{-1}$), high-throughput (≤ $10^{-17}$ erg cm$^{-2}$ s$^{-1}$ Å$^{-1}$ in $10^4$ sec)** spectroscopy would allow us to characterize the $H_2$ and CO profiles in unprecedented detail (Figure 2). This would permit the measurement of the radial gas profiles in rotating disks and the contribution from low-velocity disk winds (e.g., Pontoppidan et al. 2011) that may determine gas disk lifetimes. We emphasize that high-resolution UV spectroscopy will be the

*only* means of resolving the structures of the $H_2$ disk; the highest resolution spectroscopic modes on *JWST* are limited to ($\Delta v \geq 100$ km s$^{-1}$), therefore UV observations will be critical for interpreting *JWST* observations of $H_2$, $H_2O$, and other molecules at terrestrial planet-forming radii.

UV observations of photoexcited CO and $H_2$ emission are also the best measure of the Ly$\alpha$ radiation field incident on the disk surface (Herczeg et al. 2004; Schindhelm et al. 2012). The complete Ly$\alpha$ emission line is not directly observable due to absorption and scattering in the intervening material, but constitutes ~80% of the total UV luminosity incident on the circumstellar environment and is essential for understanding protoplanetary disk chemistry at the epoch of planet-formation (Fogel et al. 2011).

The composition and physical state (e.g., temperature, turbulent velocity, ionization state) of a cross-section of the circumstellar environment can be probed using high-resolution absorption line spectroscopy of high-inclination ($i > 60°$) disks. **Spectral coverage in the 912 – 1150 Å bandpass is particularly important** for this work as the bulk of the warm/cold $H_2$ gas is only observable at $\lambda < 1120$ Å (via the Lyman and Werner ($v'$ - 0) band systems). This work has only been possible on a small number of bright objects from protoplanetary (Roberge et al. 2001; France et al. 2012c) to debris (Roberge et al. 2000) disk systems. A systematic study of circumstellar disks across the 1 – 100 Myr timescale of giant and terrestrial planet formation holds great promise for understanding the evolution of the environments in which planets form.

### 3.2 The Lifetimes of Protoplanetary Gas Disks

Fluorescent $H_2$ spectra in the 912 – 1650 Å bandpass are sensitive to gas surface densities $\leq 10^{-6}$ g cm$^{-2}$, making these data an extremely useful probe of trace amounts of primordial circumstellar gas at $r < 10$ AU around pre-main sequence F – M stars. In cases where mid-IR CO spectra or traditional accretion diagnostics (e.g., H$\alpha$ equivalent widths) suggest that the inner gas disk has dissipated, far-UV $H_2$ observations offer unambiguous evidence for the presence of a molecular disk (Ingleby et al. 2011b; France et al. 2012b). There is growing evidence that gas-rich disks can persist to several times the 2 – 4 Myr dust dissipation timescale (Figure 3), however this work has been limited by small sample sizes in the UV spectroscopic surveys ($<10^2$) compared to the dust SEDs compiled from mid-IR photometry and spectroscopy ($>10^4$; each square in Figure 3 represents 10s - 100s of stars). Uniform spectral surveys of entire local star-forming regions are required for a systematic determination of the gas disk lifetimes and therefore the timescales for gas envelope

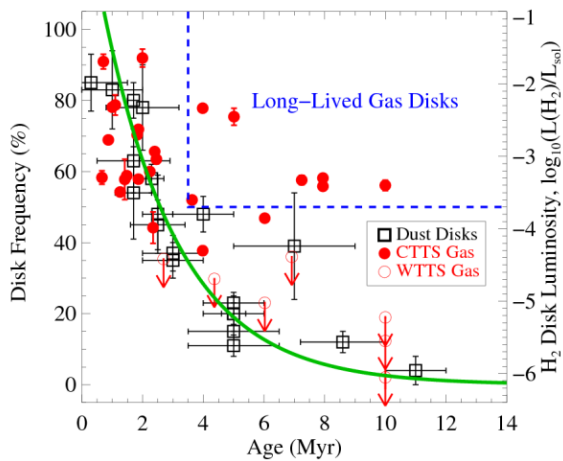

Figure 3: Far-UV $H_2$ emission lines are a sensitive measure of the molecular disk surface. Dust disk dissipation has a characteristic timescale of 2 – 4 Myr (open squares, adapted from Wyatt 2008), while a growing number of gas-rich disks are observed to persist to ≈ 4 - 10 Myr (filled red circles; France et al. 2012b).

accretion and migration of planetary cores through their natal disks.

A far-UV survey of $H_2$ and CO disks naturally lends itself to a **multi-object spectrograph (MOS, 10´ × 10´ field-of-view)** approach and is a strong science driver for the development of such an instrument for a future mission. **Moderate spectral resolution ($\Delta v \approx$ 100 km s$^{-1}$, at $F_\lambda \leq 10^{-16}$ erg cm$^{-2}$ s$^{-1}$ Å$^{-1}$ in $10^3$ sec)** is sufficient to separate blended $H_2$ emission lines for robust spectral identification and flux measurement. By studying a range of star-forming regions, from ~ 1 Myr (e.g., the Orion Nebula Cluster) to ~30 Myr (e.g., the Tucana/Horologium Association), we can not only statistically characterize the gas disk lifetimes, but also constrain the evolution of the UV radiation field (from accretion-dominated to chromosphere-dominated) incident on the terrestrial and giant planet formation regions during the growth of planetary cores and atmospheres. Combining this far-UV spectral survey with roughly contemporaneous near-UV (1700 – 4000 Å) multi-band imaging or low-resolution spectroscopy would enable the measurement of robust mass accretion rates. Comparing the mass accretion rates with the disk lifetimes will allow us to better understand the physical processes that govern the dissipation of primordial protoplanetary disks and the transition to gas-poor debris disk systems.

4. Summary

We have given four sample investigations where a future NASA mission with a UV spectroscopic capability would provide fundamentally new insight into how exoplanetary systems form and the physics that governs their atmospheres. **We argue that high-sensitivity and low background equivalent fluxes are a requirement for this mission.** Advances in component technology such as high-reflectivity UV coatings (Beasley et al. 2012; factor of 3 improvement per optic at $\lambda < 1100$Å) and low-noise borosilicate glass photon-counting detectors (Siegmund et al. 2011; factor of ~10 lower noise than *HST*-COS detectors) will provide many of the advantages of a large telescope for a fraction of the cost. We suggest that including both a high-resolution point source spectrograph and a MOS operating at lower resolution will provide the largest grasp in observatory discovery space for exoplanet and related research. Technology investment in low-scatter echelle UV spectrographs (e.g., France et al. 2012d; factor of up to ~100 improvement in scattered light control at $R > 10^5$) would provide a means for achieving the order-of-magnitude gains necessary to carry out the science without the commensurate increase in telescope diameter. We would be happy to present a summary of this report at a workshop on future UV/visible science drivers.

**References**

Armitage et al. 2002MNRAS.334..248A; Armitage 2007ApJ...665.1381A; Ben Jaffel & Hosseini 2010ApJ...709.1284B; Beasley et al. 2012SPIE8443.0B; Benneke & Seager 2012ApJ...753..100B; Blandford et al. 2010, Nat Acad Press; Ekenback et al. 2010ApJ...709..670E; Fogel et al. 2011ApJ...726...29F; Fossati et al. 2010ApJ...714L.222F; France et al. 2012aApJ...750L..32F; France et al. 2012b_arXiv1207.4789F; France et al. 2012cApJ...744...22F; France et al. 2012dSPIE8443.0B; Garcia Muñoz 2007P&SS...55.1426G; Hayashi et al. 1985prpl.conf.1100H; Herczeg et al. 2002ApJ...572..310H; Herczeg et al. 2004ApJ...607..369H; Hernández et al. 2007ApJ...662.1067H; Ingleby et al. 2011aAJ....141..127I; Ingleby et al. 2011bApJ..743..105I; Kaltenegger et al. 2011ApJ...733...35K; Koskinen et al. 2010ApJ...723..116K; Lecavelier et al. 2008A&A...481L..83L; Linsky et al. 2010ApJ...717.1291L; Murray-Clay et al. 2009ApJ...693...23M; Pontoppidan et al. 2011ApJ...733...84P; Rappaport et al. 2012ApJ...752....1R; Rauer et al. 2011A&A...529A...8R; Roberge et al. 2000ApJ...538..904R; Roberge et al. 2001ApJ...551L..97R; Salyk et al. 2009ApJ...699..330S; Schindhelm et al. 2012b_arXiV; Segura et al. 2005AsBio...5..706S; Siegmund et al. 2011SPIE.8145E.251S; Sing et al. 2011MNRAS.416.1443S; Tremblin & Chiang 2012arXiv1206.5003T; Trilling et al. 2002A&A...394..241T; Vidal-Madjar et al. 2003Natur.422..143V; Vidal-Madjar et al. 2004ApJ...604L..69V; Ward 1997Icar..126..261W; Wyatt 2008ARA&A..46..339W; Yelle 2004Icar..170..167Y